\documentclass[fleqn,10pt]{wlscirep}
\title{Magnetic Tunnel Junction Mimics Stochastic Cortical Spiking Neurons}

\author[1,*]{Abhronil Sengupta}
\author[1,+]{Priyadarshini Panda}
\author[1,+]{Parami Wijesinghe}
\author[1]{Yusung Kim}
\author[1]{Kaushik Roy}
\affil[1]{Purdue University, School of Electrical \& Computer Engineering, West Lafayette IN 47907, USA}

\affil[*]{asengup@purdue.edu}

\affil[+]{Authors contributed equally to this work}

\begin{abstract}
Brain-inspired computing architectures attempt to mimic the computations performed in the neurons and the synapses in the human brain in order to achieve its efficiency in learning and cognitive tasks. In this work, we demonstrate the mapping of the probabilistic spiking nature of pyramidal neurons in the cortex to the stochastic switching behavior of a Magnetic Tunnel Junction in presence of thermal noise. We present results to illustrate the efficiency of neuromorphic systems based on such probabilistic neurons for pattern recognition tasks in presence of lateral inhibition and homeostasis. Such stochastic MTJ neurons can also potentially provide a direct mapping to the probabilistic computing elements in Belief Networks for performing regenerative tasks.
\end{abstract}
\begin{document}

\flushbottom
\maketitle
\thispagestyle{empty}

\section*{Introduction}

The human brain is the most powerful and yet energy efficient computing system known to humans. As an attempt to mimic the human brain, and thereby emulate its efficiency in cognitive and perception tasks, computing models have been developed that try to mimic the functionalities involved in the neurons and synapses in the human brain. Although a complete understanding of the brain has still remained elusive, recent advances in neuroscience have brought forward important behavioral characteristics and phenomena underlying neuronal and synaptic operations. Neuromorphic computing refers to the emulation of such underlying neuroscience mechanisms by an equivalent hardware implementation. 

A neural network consists of neurons interconnected by synaptic junctions, which encode the importance or ``weight'' of the information transmitted by the neurons. Different abstract computing models have been developed to emulate the information processing that occurs in the biological neuron. The computing model offering the highest degree of bio-fidelity is that of the spiking neuron, which is characterized by a membrane potential that integrates incoming spikes and leaks in the absence of spikes. The neuron generates an output spike when the membrane potential crosses a specific threshold. Past research on hardware implementation of spiking neurons have mainly focused on deterministic neural models, like the Hodgkin-Huxley \cite{ghosh2009spiking} and Leaky-Integrate-Fire \cite{ghosh2009spiking} models. However, emulation of such neural characteristics require area-expensive CMOS implementations involving more than 20 transistors \cite{rajendran2013specifications,indiveri2003low} and a direct mapping of spiking neuronal characteristics to a single nanoelectronic device is still missing. Further, such deterministic neuron models have little correspondence to the probabilistic firing nature of biological neurons and are unable to account for the fact that neural computation in the brain is significantly prone to noise arising from the synapses, dendrites or the neuron itself \cite{sobie2011neuron,nessler2013bayesian}. 

Recently, theoretical studies have been performed to demonstrate that Bayesian computation can be performed in networks inspired from cortical microcircuits of pyramidal ``stochastic" neurons \cite{nessler2013bayesian}. Such neurons, observed in the cortex, spike stochastically and the probability of firing at a particular time is a non-linear function of the instantaneous magnitude of the resultant post-synaptic current input to the neuron \cite{nessler2013bayesian,wallace2011emergent,benayoun2010avalanches,nessler2009stdp}. In this paper, we demonstrate a nano-magnetic device that can mimic such cortical ``stochastic" spiking neurons. 

\section*{Magnetic Tunnel Junction as a spiking neuron}

\begin{figure}[!t]
\centering
\includegraphics[width=2.6in]{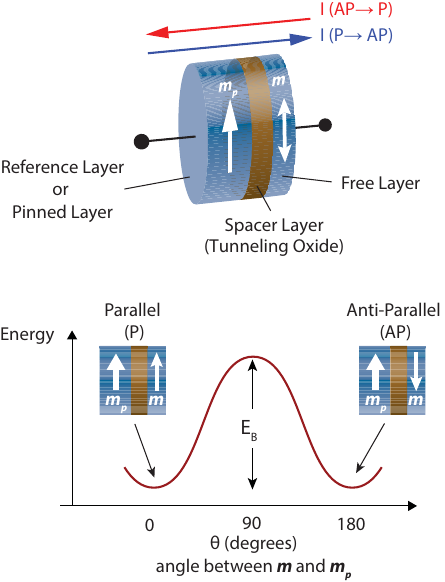}
\caption{\footnotesize{A Magnetic Tunnel Junction (MTJ) consists of two magnetic layers sandwiching a spacer layer. While the magnetization direction of the reference layer is pinned, the magnetization of the free layer can be manipulated by an input charge current. The MTJ is characterized by two stable resistance states, namely the parallel (P) and anti-parallel (AP) configuration. The barrier height ($E_{B}$) causes the P and AP states of the MTJ to be thermally stable.}}
\label{mtj}
\end{figure}

Let us first illustrate the device structure and principle of operation of a Magnetic Tunnel Junction (MTJ) \cite{julliere1975tunneling,baibich1988giant,binasch1989enhanced}. The MTJ consists of two ferromagnetic layers separated by a tunneling oxide barrier ($MgO$). The magnetization direction of one of the layers (denoted by $pinned$ layer, PL, in Fig. \ref{mtj}), $\widehat {\textbf {m}}_P $, is magnetically hardened so that it serves as the reference layer. The magnetization of the $free$ layer (FL), $\widehat {\textbf {m}} $, can be manipulated by an input charge current. The MTJ is characterized by two stable resistance states, namely the low-resistance parallel (P) configuration ($\widehat {\textbf {m}} $ and $\widehat {\textbf {m}}_P$ are parallel) and the high-resistance anti-parallel (AP) configuration ($\widehat {\textbf {m}} $ and $\widehat {\textbf {m}}_P$ are anti-parallel). Charge current from the $pinned$ layer to the $free$ layer causes the MTJ to switch to the AP state and vice versa by overcoming the energy barrier, $E_{B}$ (see Fig. \ref{mtj}). Considering the initial state of the MTJ to be the P state, such a behavior can be mapped to a neural firing when the MTJ switches to the AP state.

The magnetization dynamics of the FL in a nanoscale monodomain magnet at $T=0K$ can be described by solving Landau-Lifshitz-Gilbert equation with additional term to account for the spin momentum torque according to Slonczewski\cite{slonczewski1989conductance},
\begin{equation}
\label{llg}
\frac {d\widehat {\textbf {m}}} {dt} = -\gamma(\widehat {\textbf {m}} \times \textbf {H}_{eff})+ \alpha (\widehat {\textbf {m}} \times \frac {d\widehat {\textbf {m}}} {dt})+\frac{1}{qN_{s}} (\widehat {\textbf {m}} \times \textbf {I}_s \times \widehat {\textbf {m}})
\end{equation}
where, $\widehat {\textbf {m}}$ is the unit vector of free layer magnetization, $\gamma= \frac {2 \mu _B \mu_0} {\hbar}$ is the gyromagnetic ratio for electron, $\alpha$ is Gilbert \textquoteright s damping ratio, $\textbf{H}_{eff}$ is the effective magnetic field including the shape anisotropy field for elliptic disks calculated using \cite{beleggia2005demagnetization}, $N_s=\frac{M_{s}V}{\mu_B}$ is the number of spins in free layer of volume $V$ ($M_{s}$ is saturation magnetization and $\mu_{B}$ is Bohr magneton), and $I_{s}$ is the input spin current generated by charge current flow through the $pinned$ layer. Equation ~\ref{llg} can be reformulated by simple algebraic manipulations as,
\begin{equation}
\frac{1+\alpha^2}{\gamma} \frac {d\widehat {\textbf {m}}} {dt} =-(\widehat {\textbf {m}} \times \textbf {H}_{eff})- \alpha (\widehat {\textbf {m}} \times \widehat {\textbf {m}} \times \textbf {H}_{eff})+\frac{1}{q\gamma N_{s}} (\alpha(\widehat {\textbf {m}} \times \textbf {I}_s )-(\widehat {\textbf {m}} \times \widehat {\textbf {m}} \times \textbf {I}_s) )
\end{equation}
\begin{figure}[!t]
\centering
\includegraphics[width=4.4in]{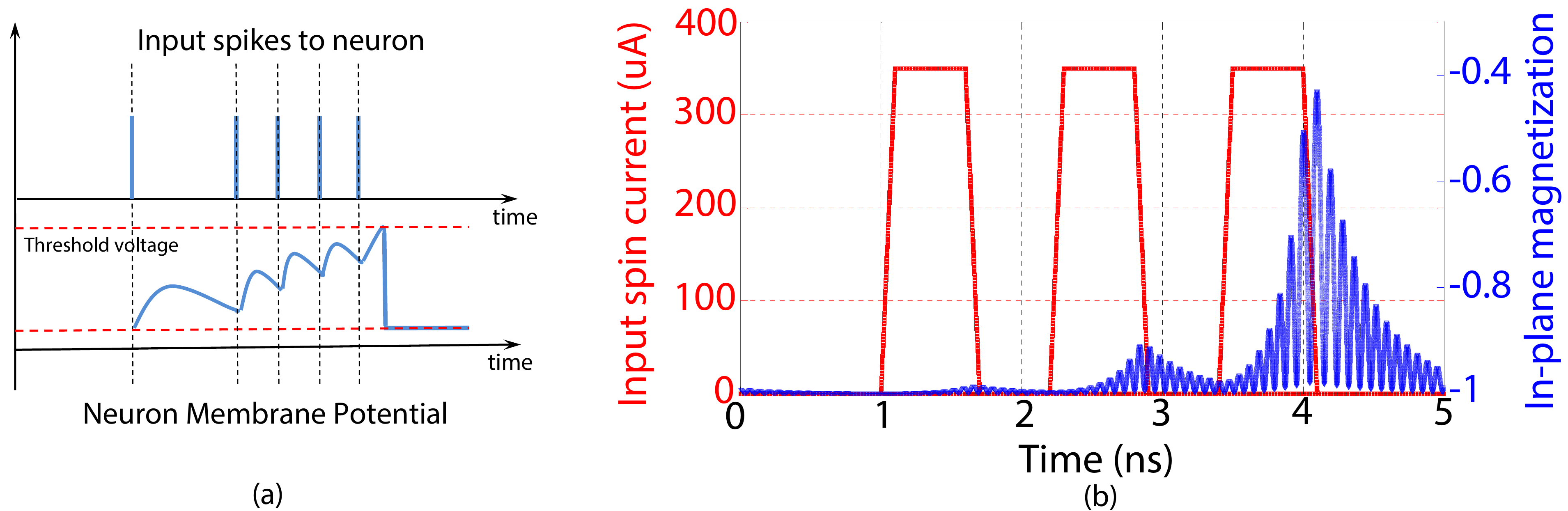}
\caption{\footnotesize{(a) The membrane potential of a biological neuron integrates input spikes and leaks when there is no input. It spikes when the membrane potential crosses the threshold. (b) MTJ neuron dynamics due to the application of three input pulses. The in-plane magnetization starts integrating due to the pulses and then starts leaking once the pulse is removed. The MTJ structure was an elliptic disk of volume $\frac{\pi}{4} \times 100 \times 40 \times 1.5 nm ^3$ with saturation magnetization of $M_s = 1000 KA/m$ and damping factor, $\alpha = 0.0122$.}}
\label{mtj_lif}
\end{figure}
Let us consider an MTJ with in-plane magnetic anisotropy (IMA). The in-plane component of magnetization, $\widehat {\textbf {m}}$, of the nanomagnet can be considered equivalent to the membrane potential of a biological neuron. The first two terms in the RHS of the above equation constitute the ``leak'' term in the magnetization (membrane potential) dynamics while the last term relates to the integration of input pulses applied to the MTJ. The MTJ ``fires'' when the magnetization switches to the opposite stable state. Fig. \ref{mtj_lif} illustrates the leak and integration components of the neuron dynamics for an MTJ elliptic disk due to the application of three successive pulses. The magnetization starts increasing due to integration of the pulses. However, it is insufficient to ``switch'' the MTJ and the magnetization starts leaking once the applied pulse is removed. The firing or ``spiking'' of the neuron (which occurs when the membrane potential crosses the threshold) is equivalent to the switching of the MTJ, i.e. magnetization reversal of the in-plane component from -1 to +1. Once the neuron ``spikes'', it has to be reset back to the initial state. Hence, the operation of the neuron MTJ can be resolved into two cycles, namely a ``write'' phase followed by a ``read'' phase. During the ``write'' phase, the MTJ neuron receives the resultant input synaptic current at a particular time step while the ``read'' phase is utilized to determine whether the neuron has switched during the ``write'' phase and is reset back to the P state in case the MTJ switched to the AP state. This reset phase is analogous to the ``refractory" period observed in biological neurons \cite{ghosh2009spiking} where the neuron is not able to generate a ``spike'' for some time duration after generating a ``spike'' (corresponding to the time delay involved in resetting back the MTJ to the P state).

\begin{figure}[!t]
\centering
\includegraphics[width=3.3in]{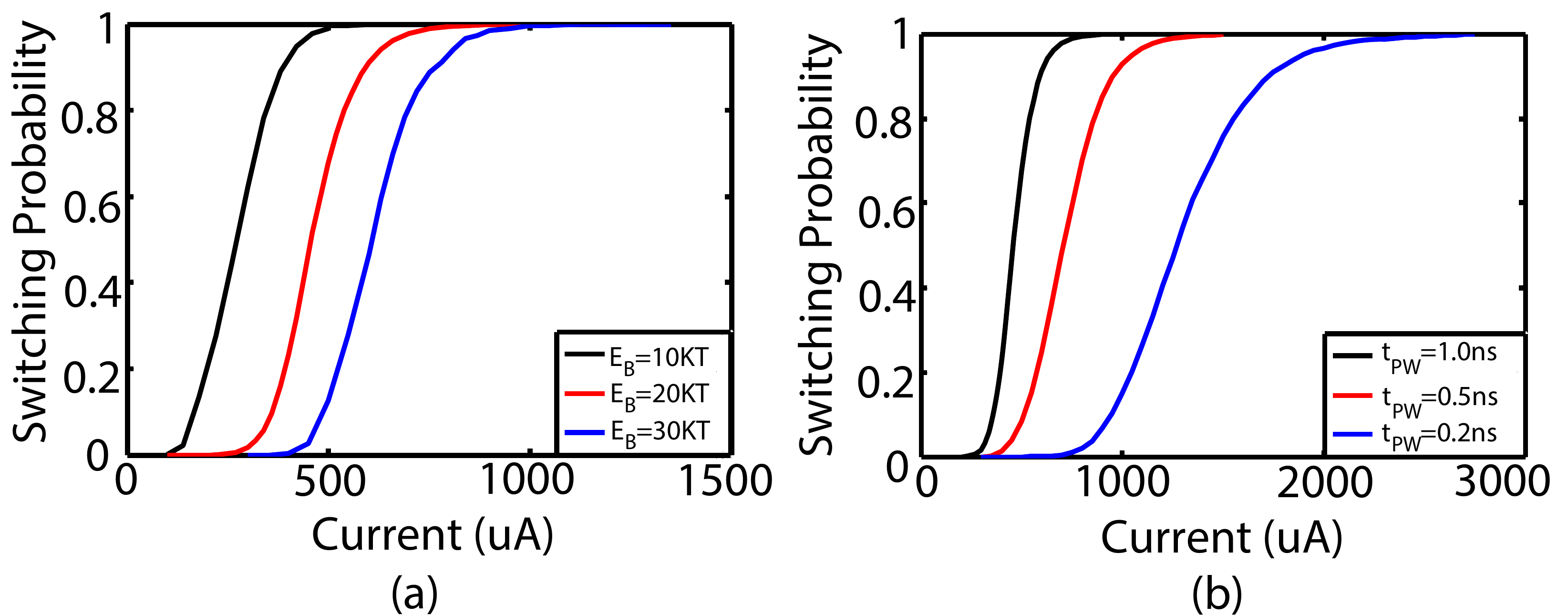}
\caption{\footnotesize{Switching probability of the MTJ in response to an input synaptic current at $T=300K$ (assuming $\sim 50 \%$ polarization of spin current generated by the MTJ $pinned$ layer). Such a switching behavior is a direct mapping to the stochastic spiking nature of cortical neurons. (a) The switching probability characteristics shifts to the right with increase in the barrier height. The data have been plotted for $E_{B} = (10, 20, 30) K_{B}T$ corresponding to FL thickness values, $t_{FL} = (0.8, 1.2, 1.5) nm$, for pulse width, $t_{PW}=1ns$ (duration of the ``write'' cycle). (b) The probability characteristics undergo more dispersion with decrease in the pulse width. The data have been plotted for $t_{PW} = (0.2, 0.5, 1) ns$ corresponding to $E_{B} = 20K_{B}T$.}}
\label{prob}
\end{figure}

\begin{figure*}[!t]
\centering
\includegraphics[width=6.6in]{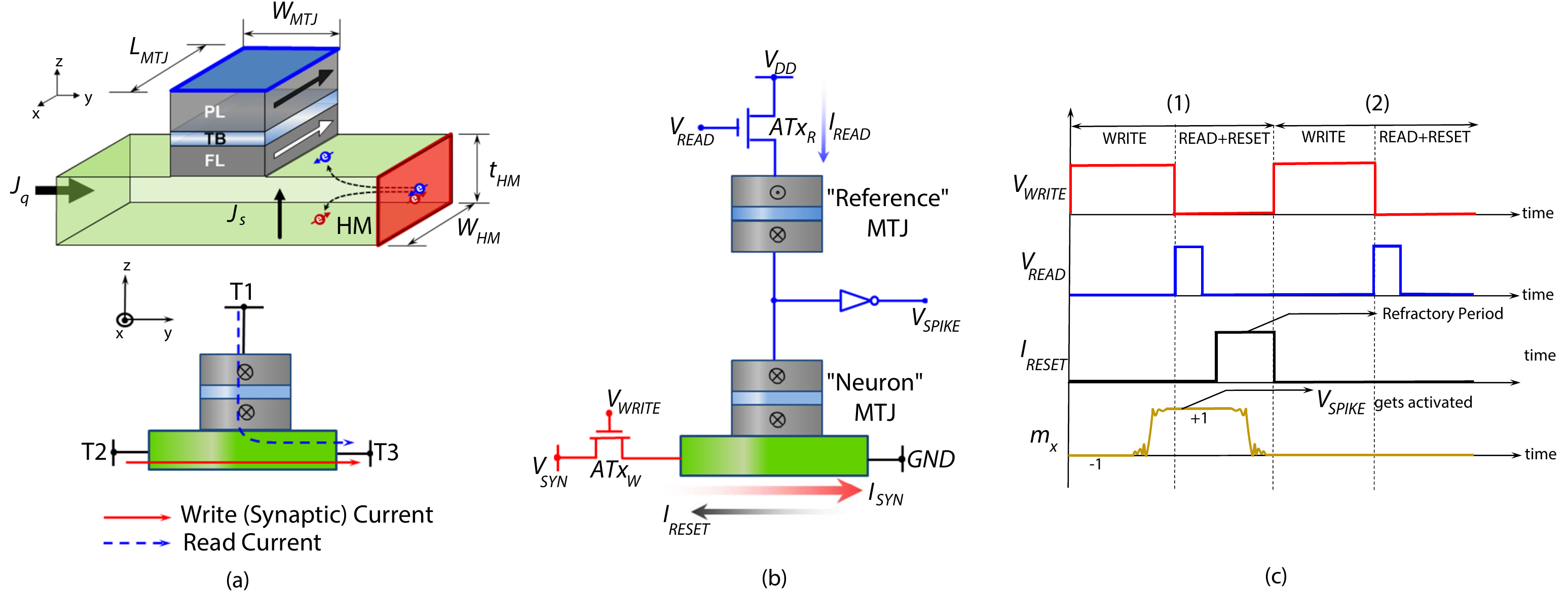}
\caption{\footnotesize{(a) Input charge current flowing in the +y direction through a heavy-metal (HM) with high-spin orbit coupling causes accumulation of +x directed spins at the interface of the HM and a ferromagnet ``free layer" (FL) lying at the top. The FL with in-plane uniaxial anisotropy (IMA) can be switched by the current flowing through the HM. Schematic of the three-terminal device proposed as a stochastic neuron with decoupled ``read" and ``write" current paths. The input synaptic current flows between terminals T2 and T3 while the read current flows through T1 and T3. (b) The stochastic neuron (``Neuron MTJ") is interfaced with access transistors to decouple the ``write" and ``read" current paths. During the ``write" cycle ($V_{WRITE}$ activated), the incoming synaptic current, $I_{SYN}$, in presence of thermal noise, probabilistically switches the neuron depending on its magnitude. During the subsequent ``read" cycle ($V_{READ}$ activated), a small current $I_{READ}$ flows through the two MTJs in series. The ``Reference" MTJ's magnetization is fixed to the AP state causing the inverter to generate a spike ($V_{SPIKE}$) in case the neuron switches from the P to the AP state. In case the neuron spiked, the neuron is reset to the P state using a reset current $I_{RESET}$. The peripheral circuit for resetting the neuron involves a similar access transistor connecting the device to a ``reset'' voltage, whose gate is driven by the output of a latch that stores the value of the spike signal, $V_{SPIKE}$, at the end of the``read'' cycle.(c) Two complete periods are shown to explain the operation in detail.}}
\label{she}
\end{figure*}

At non-zero temperature, the magnetization dynamics of the MTJ is characterized by thermal noise, which can be accounted for by an additional thermal field \cite{scholz2001micromagnetic}, $\textbf{H}_{thermal}=\sqrt{\frac{\alpha}{1+\alpha^{2}}\frac{2K_{B}T}{\gamma\mu_{0}M_{s}V\delta_{t}}}G_{0,1}$, where $G_{0,1}$ is a Gaussian distribution with zero mean and unit standard deviation, $K_{B}$ is Boltzmann constant, $T$ is the temperature and $\delta_{t}$ is the simulation time step. In presence of thermal noise, the switching behavior of the MTJ due to the flow of a charge current through the $pinned$ layer, during the ``write'' cycle, is stochastic in nature and the probability of switching increases with increase in the magnitude of input current. Hence, such a device offers a direct mapping to the functionality of ``stochastic" neurons observed in the cortex \cite{nessler2013bayesian,wallace2011emergent,benayoun2010avalanches,nessler2009stdp}, where the neuron ``spikes" (switches its state) probabilistically depending on its resultant synaptic input. The variation of spiking probability with input synaptic current is usually described by a non-linear dependence \cite{nessler2013bayesian,wallace2011emergent,benayoun2010avalanches,nessler2009stdp}, similar to the MTJ switching characteristics shown in Fig. \ref{prob}. The switching characteristics of the MTJ neuron in response to the input synaptic current can be varied by changing the energy barrier (or equivalently the $free$ layer thickness) and the duration of the synaptic current as illustrated in Fig. \ref{prob}. 

Recent experiments have shown that such an MTJ structure with in-plane magnetic anisotropy (IMA) can also be switched by a charge current flowing through a heavy-metal (HM) underlayer due to the injection of spins (whose polarization is transverse to the direction of both spin and charge current) at the FL-HM interface (assuming spin-Hall effect to be the dominant underlying physical phenomenon: Fig. \ref{she}(a)) \cite{hirsch1999spin,pai2012spin,miron2011perpendicular,liu2012spin,liu2012current}. We will refer to FL switching by such a HM underlayer for the rest of this text due to the possibilities of having decoupled ``write'' and ``read'' current paths which helps in interfacing such MTJ ``stochastic'' neurons with a synaptic resistive crossbar array (discussed later in the text). It is worth noting here that the mechanism of MTJ switching by spin-Hall effect and mapping to a neuron functionality is exactly similar as discussed before. The only difference is that the spin current is generated by the HM underlayer instead of the $pinned$ layer of the MTJ. The generated spin current, $I_{s}=\theta_{SH}\frac{W_{MTJ}}{t_{HM}}I_{Q}$ ($I_{Q}$ is the charge current flowing through the HM, $\theta_{SH}$ is spin-Hall angle \cite{pai2012spin}, dimensions $W_{MTJ}$ and $t_{HM}$ are shown in Fig. \ref{she}(a)). Hence, the device also offers energy-efficient ``write" since spin polarization is not limited by polarization of the $pinned$ layer and $>100\%$ spin injection efficiency can be achieved \cite{pai2012spin}. The device simulation parameters were obtained from experimental measurements \cite{pai2012spin} and have been shown in Table I. Fig. \ref{she}(b)-(c) illustrates the principle of operation of the ``Neuron" MTJ with access transistors to decouple the ``write" and ``read" current paths.
\begin{figure*}[!t]
\centering
\includegraphics[width=6.6in]{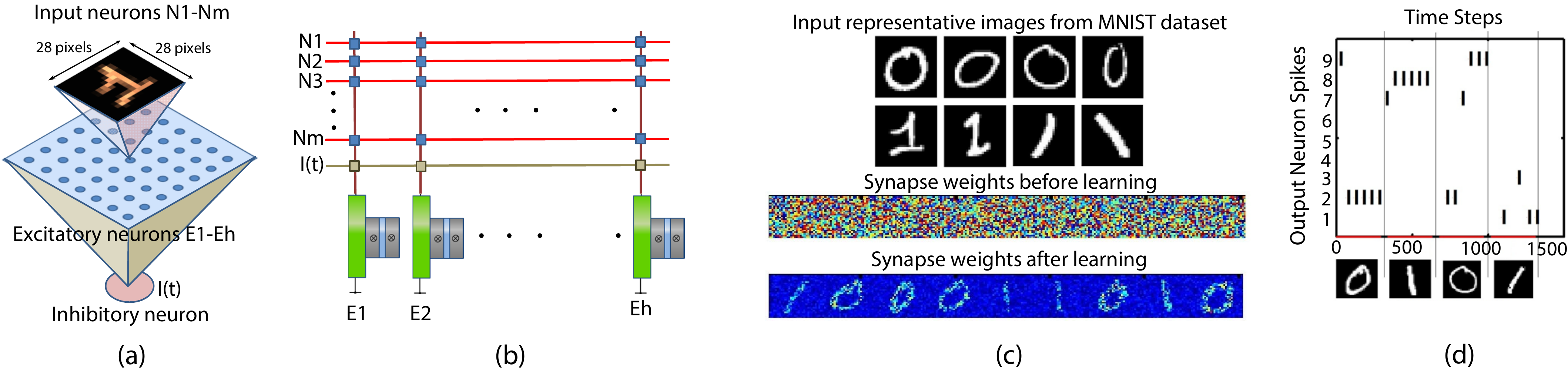}
\caption{\footnotesize{(a) Stochastic spiking neural network used for digit recognition. Input spike trains are received by all the stochastic neurons (connections shown for only one neuron). The inhibitory neuron prevents the neurons from spiking in case an excitatory neuron spikes. (b) Corresponding implementation in a crossbar array fashion. Programmable resistive synapses are present at each cross-point. Input voltages are applied at each row and the neurons receive input synaptic current which is the weighted summation of the input voltages. (c) A network of 9 excitatory neurons were used for the recognition purpose. The synapse weights were randomly initialized. 784 input neurons (28 x 28 images) are rate encoded by ensuring that the spike frequency is directly proportional to the pixel intensity. After learning the neurons respond selectively to each input image. (d) For testing the behavior of the network after learning has been accomplished, STDP and homeostasis were turned off. The neuron stochastically spikes the maximum for the class which it has learnt while the others remain mostly silent. A common lateral inhibitory signal during testing results in sparse spiking events.}}
\label{network}
\end{figure*}
\begin{table}[h]
\label{table}
\center
\centerline{TABLE I. Device Simulation Parameters}
\vspace{2mm}
\begin{tabular}{c c}
\hline \hline
\bfseries Parameters & \bfseries Value\\
\hline
Free layer area & $\frac{\pi}{4} \times 100 \times 40 nm^2$\\
Free layer thickness & $ 0.8, 1.2, 1.5 nm$\\
Heavy-metal thickness, $t_{HM}$ & $ 2 nm$\\
Saturation Magnetization, $M_{S}$ & 1000 $KA/m$ \cite{pai2012spin}\\
Spin-Hall Angle, $\theta_{SH}$ & 0.3 \cite{pai2012spin} \\
Gilbert Damping Factor, $\alpha$ & 0.0122 \cite{pai2012spin} \\
Energy Barrier, $E_{B}$ & 10, 20, 30 $K_{B}T$ \\
MgO Thickness, $t_{MgO}$ & 2$nm$ \\
MTJ Resistance in P (AP) state, $R_P(R_{AP})$ & 1.21 (2.5) $M\Omega$ \\
Resistivity of HM, $\rho_{HM}$ & 200 $\mu\Omega.cm$ \cite{pai2012spin}\\
Pulse width, $t_{PW}$ & $0.2, 0.5, 1ns$ \\
Temperature, $T$ & $300K$ \\
Supply Voltage, $V_{DD}$ & $1V$ \\
\hline \hline
\end{tabular}\\ 
\end{table}

\section*{Spiking Neural Network based on MTJ neurons}

The behavior of a network of such stochastic MTJ neurons were studied in a standard digit recognition problem based on the MNIST dataset \cite{lecun1998gradient} as shown in Fig. \ref{network}(a). Such network connections have been observed in pyramidal neurons in the cortex \cite{nessler2013bayesian,nessler2009stdp}. The neurons receive input Poisson spike trains whose frequency is proportional to the pixel intensity. 100 images of digits ``0" and ``1" were used for the recognition purpose and the network was simulated for a number of time steps, $T_{S}$, for each image. It is worth noting here that each time step refers to the duration of the ``write'' phase of the neuron MTJ discussed before. Whenever a neuron spikes, a common inhibitory signal prohibits the neurons from spiking for a period, $\tau_{inh}$. Hence, during learning, lateral inhibition prevents the non-spiking neurons from spiking for a particular duration, thereby causing the spiking neurons to start responding selectively to specific input patterns. However, in order to prevent single neurons from dominating the spiking pattern, homeostasis \cite{diehl2015unsupervised,knag2015sparse} is performed by scaling the input current to the MTJ neuron by a variable which increases as learning progresses. Interested readers are referred to Ref. \cite{diehl2015unsupervised} for a detailed description of pattern recognition performed in such spiking networks enabled by lateral inhibition and homeostasis. Such a network arrangement can be mapped to a crossbar network interfaced with such MTJ neurons as shown in Fig. \ref{network}(b) where programmable resistive synapses encode the synaptic weight at each cross-point. Phase-change devices \cite{jackson2013nanoscale}, Ag-Si memristors \cite{jo2010nanoscale} or spintronic synapses \cite{sengupta2015spin} have been proposed in literature to implement such synaptic functionality in a crossbar architecture. The synapses were modeled with 4-bit discretization and a maximum to minimum resistance ratio of 20. An input spike triggers a voltage across the corresponding row for a duration of $\tau_{0}$ time steps (analogous to post-synaptic potential observed in biology). The neuron, therefore, receives an input current which is proportional to the weighted sum of the post-synaptic voltages (since HM resistance is much lower than the synaptic resistances at each cross-point) and spikes in a stochastic manner. A behavioral model of the neuron was developed by running stochastic LLG simulations to capture its probabilistic spiking behavior. Non-Equilibrium Green's function based transport simulation framework \cite{fong2011knack} was used to model the MTJ resistance. Unsupervised learning was performed using Spike-Timing Dependent Plasticity (STDP) \cite{diehl2015unsupervised,knag2015sparse}. The STDP weight update equations were, $\Delta w = \eta_{+} w \exp (\frac{-\Delta t}{\tau_{+}})$ (for $\Delta t >0$) and $\Delta w = -\eta_{-} w \exp (\frac{\Delta t}{\tau_{-}})$ (for $\Delta t <0$), where $\Delta t$ is the spike timing difference. The neurons learn representative models of the digits after a few epochs (Fig. \ref{network}(c)). After learning, each neuron gets trained to respond to a specific digit (Fig. \ref{network}(d)). Such learning functionalities can be exploited to develop pattern recognition systems where the input image class is detected from the spiking patterns of the neurons in the network. The network simulation parameters have been outlined in Table II. 

\begin{figure}[!t]
\centering
\includegraphics[width=3.4in]{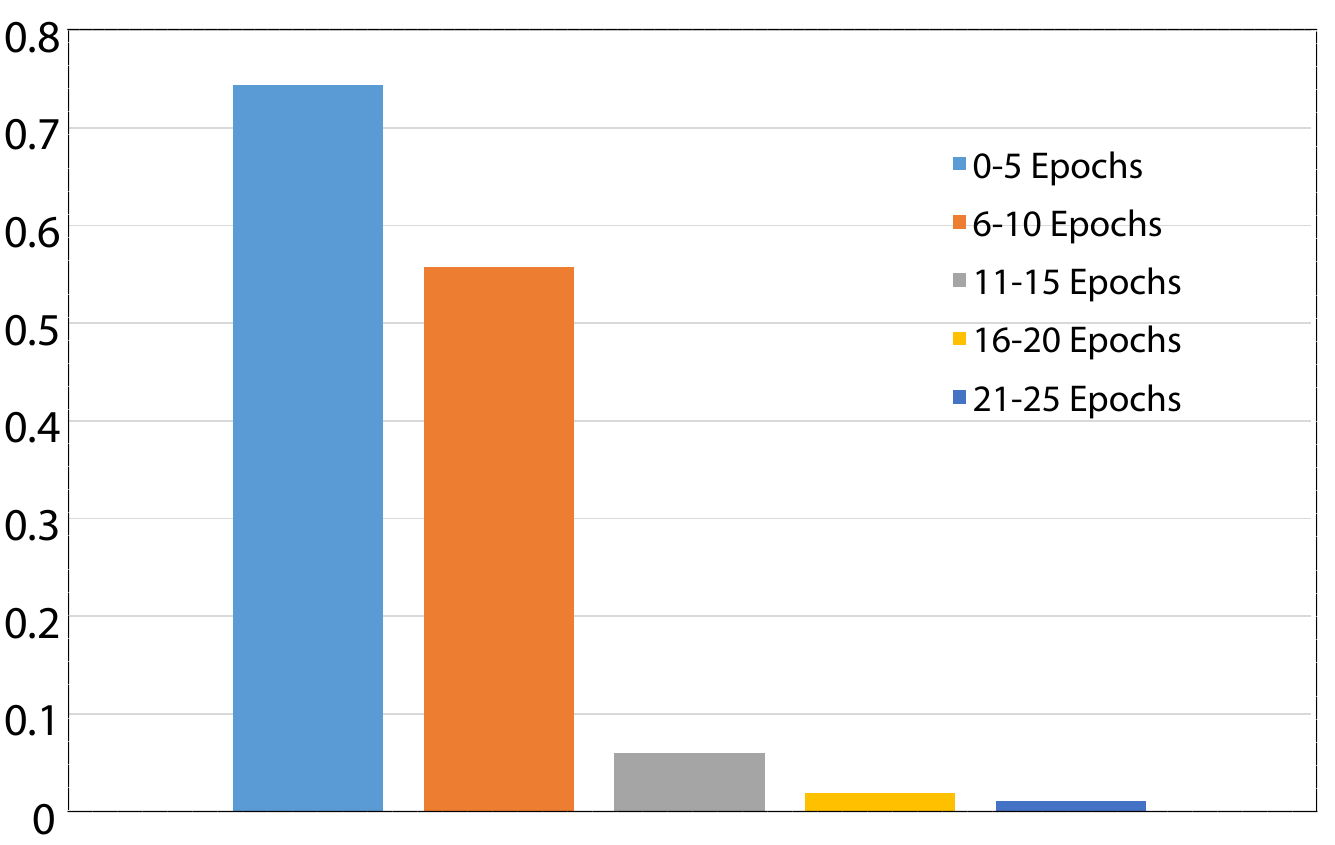}
\caption{\footnotesize{Variation of the MTJ switching probability during the learning process. While the switching probability is high during the initial learning process, it gradually converges to low values due to homeostasis.}}
\label{prob_vs_epoch}
\end{figure}
Fig. \ref{prob_vs_epoch} illustrates the manner in which the entire switching probability characteristics of the MTJ (from the deterministic to the stochastic regime) is exploited to realize learning functionality. The figure represents the maximum switching probability of the MTJ in the network (representing the neuron with the highest spiking activity which attempts to learn an applied input pattern) averaged over a range of 5 learning epochs from the beginning of the learning process. Each epoch corresponds to the entire duration of spike-train applied as input for a particular image. As explained previously, initially the switching probability of the MTJ is sufficiently high in order to ensure that different neurons start learning different input patterns. However, as learning progresses, due to homeostasis, the spiking probability of the neurons reduce resulting in sparser neural and learning events. Readers are referred to Ref. \cite{nessler2013bayesian} for an extensive theoretical discussion on probabilistic Bayesian computation that can be performed using such stochastic spiking neurons in networks inspired from cortical connections. The efficiency of the learning process can be observed from Fig. \ref{network}(d) where a slight variation in the orientation of digits belonging to the same class can be detected in the spiking activities of the neurons.  

Let us now provide a brief discussion on the energy efficiency of the system. Each ``write'' and ``reset'' cycle for a particular time step in the simulation was taken to be $0.5ns$ long and the barrier height of the neuron MTJ was chosen to be $20K_{B}T$. The energy consumption of the neuron during the ``write'' cycle is a function of the input synaptic current. Since the entire switching probability characteristics is exploited during the learning process, the average energy consumption was determined for the input current ($\sim 71\mu A$) necessary to switch the MTJ with a probability of 0.5. The associated $I^{2}Rt$ ``write'' energy consumption was evaluated to be $\sim 1fJ$ per neuron per time-step. Circuit simulations of the ``read'' circuit, shown in Fig. \ref{she}(b), yielded an average energy consumption of $\sim 1.6fJ$ per neuron per time-step (including the resistive divider circuit and the inverter). Additionally the neuron can be reset by passing a high enough reset current through the HM in the opposite direction to ensure deterministic MTJ switching. Assuming a reset current of $150 \mu A$, the $I^{2}Rt$ ``reset'' energy consumption is evaluated to be $\sim 4.5fJ$. Note that ``reset'' energy consumption is only involved in the neuron in the case of a spiking event. Additionally, the energy consumption involved in clocking the ``write'' and ``read'' cycles per time-step would result in insignificant contribution to the total energy consumption per neuron since it can be achieved by a global control circuit for the entire network of neurons. In contrast, state-of-the-art designs of CMOS neurons result in energy consumption in the range of $pJ$ per spike ($267pJ$ reported in Ref. \cite{livi2009current} and $41.3pJ$ reported in Ref. \cite{joubert2012hardware}). The energy and area benefits offered by networks of such stochastically spiking MTJ neurons in comparison to conventional deterministic spiking neuron designs in CMOS technology is the main motivation behind this proposal.
\begin{table}[h]
\label{table}
\center
\centerline{TABLE II. Network Simulation Parameters}
\vspace{2mm}
\begin{tabular}{c c}
\hline \hline
\bfseries Parameters & \bfseries Value\\
\hline
Probability of input spikes per time-step & $0-0.064$\\
Number of time-steps per image, $T_{S}$ & $340$\\
Post-synaptic voltage duration, $\tau_{0}$ & $50^1$\\
Inhibitory signal duration, $\tau_{inh}$ & $50^1$ \\
STDP time constants, $\tau_{+}, \tau_{-}$ & $4.5, 5^1$ \\
STDP learning rates, $\eta_{+}, \eta_{-}$ & $0.03, 0.01$ \\
Duration of each time-step & 0.5ns \\
Maximum synapse resistance in crossbar array & $3.7M\Omega$ \\
HM resistance & 400$\Omega$ \\
\hline \hline
\end{tabular}\\
\vspace{3mm}
$^1$ The units are in terms of time-step (i.e. $0.5ns$).
\end{table}

\section*{Conclusions}

To conclude, researchers have explored MTJs as synapses \cite{thomas2015tunnel,krzysteczko2012memristive,locatelli2014spin,vincent2014spin} and for inter-neuron communication \cite{krzysteczko2012memristive} previously. Further, previous research on utilizing spintronic devices as neurons \cite{sharad2012spin,:/content/aip/journal/apl/106/14/10.1063/1.4917011} have been limited to emulating only thresholding operations of non-spiking neural computing models. On the other hand, spiking neurons offer a more biologically realistic perspective and are recently becoming popular computing models for implementing low-power, high accuracy recognition platforms in complex cognitive tasks \cite{diehl2015fast}. 

Stochasticity exhibited by phase change memory \cite{suri2012cbram} and spintronic devices \cite{vincent2014spin} have been exploited previously in neuromorphic applications to implement learning functionality in synapses. However, the utilization of device stochasticity in nanoelectronic neural computing has been a relatively unexplored area. To the best of our knowledge, this is the first demonstration of mapping the stochastic leaky-integrate switching behavior of MTJs in presence of thermal noise to a probabilistic spiking neuron. An important point worth considering is whether other post-CMOS technologies \cite{suri2012cbram} exhibiting stochastic switching characteristics could be potentially operated as neurons as well. A few words regarding the architecture of the pattern recognition system (Fig. \ref{network}) are in order to outline the prospective opportunities offered by spintronic neurons. Neurons need to be interfaced with a crossbar array of resistive synapses for any pattern recognition system. Memristive devices are present at each cross-point to encode the synaptic weight. Input voltages are applied across each row and the current flowing through the memristors is weighted by its conductance and gets summed up along the column and passes as input to the neuron. However, this is true only when the input resistance of the neuron is sufficiently low since otherwise, the voltage drop across each memristor will be dependent on the voltage drop across the neuron which in turn, depends on the total amount of input synaptic current resulting in a coupled system. Low terminal voltage of MTJ neurons during ``write'' operation offers unique possibilities in this regard. Input synaptic current flows through the HM (with low resistance) and not through the oxide layer of the MTJ. Thus decoupled ``read'' and ``write'' current paths of the proposed neuron  assist the neuron operation. In contrast, memristive devices are usually characterized by high threshold voltages ($>1V$) and high resistance values ($K\Omega$-$M\Omega$ \cite{jackson2013nanoscale,jo2010nanoscale,suri2012cbram}). Hence, although intrinsic noise might be present in memristive devices, it will be potentially difficult to interface memristive synaptic crossbar arrays with memristive neurons.

Although the impact of thermal noise on MTJ switching behavior has limited its scalability in memory applications, such noise effects can be potentially exploited to build probabilistic neural computing platforms that can perform Bayesian computation similar to the brain. Past research on hardware implementation of spiking neurons has mainly focused on the emulation of deterministic spiking neural characteristics and require area and power expensive CMOS implementations involving more than 20 transistors \cite{indiveri2003low, rajendran2013specifications}. CMOS based stochastic neural models might be possible \cite{alspector1989performance} but involve significant silicon area and power consumption since they do not offer a direct mapping to the underlying neuroscience mechanisms. However, the ultra-low current induced noisy switching characteristics of MTJs can efficiently mimic such stochastic spiking neural models and can potentially pave the way for neuromorphic systems that utilize noisy stochastic neurons as a computing element, such as Restricted Boltzmann Machines and Deep Belief Networks. We would like to conclude the paper by noting that the device stochasticity observed in such MTJ structures can be utilized to realize probabilistic learning functionality in single-bit synapses \cite{vincent2014spin} which could be potentially interfaced with stochastic MTJ neurons resulting in an All-Spin neuromorphic architecture that leverages the underlying device stochasticity to perform neuromimetic computing.

\section*{Acknowledgements}

The work was supported in part by, Center for Spintronic Materials, Interfaces, and Novel Architectures (C-SPIN), a MARCO and DARPA sponsored StarNet center, by the Semiconductor Research Corporation, the National Science Foundation, Intel Corporation and by the National Security Science and Engineering Faculty Fellowship.

\section*{Author contributions statement}

A. Sengupta and K. Roy conceived the research study. A. Sengupta wrote the paper, developed the simulation framework and performed the simulations. Authors P. Panda and P. Wijesinghe contributed equally to this work and assisted in performing the simulations. Y. Kim helped in developing the device simulation framework. K. Roy helped with the writing of the paper, developing the concepts and discussing the results.

\section*{Competing Financial Interests}

The authors declare no competing financial interests.

\end{document}